\documentclass[aps,prl,groupedaddress,showpacs,reprint]{revtex4-2}
\usepackage{graphicx,color}
\usepackage{amsmath}
\usepackage{amsfonts}
\usepackage{bm}
\usepackage{ulem}

\begin{document}

\title{Universal Thermodynamic Law Governing Stochastic Pendulum Clocks}

\author{Yuki Izumida}
\affiliation{Department of Complexity Science and Engineering, Graduate School of Frontier Sciences, The University of Tokyo, Kashiwa 277-8561, Japan}
\thanks{izumida@k.u-tokyo.ac.jp}

\begin{abstract}
Clocks' performance, especially their precision, is fundamentally limited by thermodynamic laws as they are physical devices.
Yet, it is known that the established thermodynamic uncertainty relation (TUR), which imposes an upper bound on the uncertainty product of the precision of oscillations and entropy production, is violated for underdamped systems such as stochastic pendulum clocks.
Here, we show that for a general class of stochastic pendulum clocks described as a weakly nonlinear oscillator the uncertainty product of the phase of a pendulum clock and entropy production takes a universal and simple form that depends solely on the degree of nonlinearity.
Its validity and limitations are examined using several representative models of pendulum clocks.
Our framework reveals a universal thermodynamic principle governing stochastic pendulum clocks beyond the conventional TUR and provides a foundation for designing optimal pendulum clocks that operate efficiently in stochastic environments.
\end{abstract}

\maketitle

{\sl Introduction}--.
Precise measurement of time is the foundation of physics.
Time is measured using clocks, which are physical devices.
Clocks' operations, from classical to quantum, are inevitably subject to noise or external disturbances and their precision are ultimately governed by thermodynamic laws~\cite{M2020,EMSWB2017,PGELBHA2021,MMSAEG2025}.

As the size of a clock decreases, the effect of noise becomes dominant.
The mechanism of timekeeping by a small-sized clock is more nontrivial compared to that by deterministic clocks.
Stochastic thermodynamics, which has rapidly developed over the past three decades, provides a powerful framework for exploring the thermodynamic laws in such a stochastic environment~\cite{US2025}.
One such law is the thermodynamic uncertainty relation (TUR)~\cite{HG2020,BS2015,GHPE2016,PRS2017,DS2018,HV2019,HV2019_2,LGU2020,KS2020}, which is usually formulated as
\begin{align}
Q_t \equiv \frac{{\rm Var}J(t)}{\left<J(t)\right>^2} \sigma_{\rm irr}t \ge 2.\label{eq.TUR}
\end{align}
Here, $J(t)$ denotes a fluctuating time-integrated current of a nonequilibrium system at time $t$, $\left<\cdot \right>$ denotes an ensemble average, 
${\rm Var}J(t)\equiv \left<J(t)^2\right>-\left<J(t)\right>^2$ denotes the variance, and $\sigma_{\rm irr}$ denotes the entropy production rate.
The product of the relative fluctuation of a current and the entropy production, which we call the {\it uncertainty product} $Q_t$, is bounded below by $2$, showing a trade-off relation between them.
The TUR~\eqref{eq.TUR} and similar trade-off relations have been widely applied to investigate the performance of stochastic oscillations such as chemical and biochemical oscillations~\cite{BS2016,KH2021,CL2024,MCH2019,YI2021,OSB2022,SF2025,NSB2018,CWQT2015,FCQT2018,CJH2020,NI2025,K2025}.

The TUR in the form of Eq.~\eqref{eq.TUR}, however, is applied only for overdamped systems without velocity degrees of freedom; it can be violated for underdamped systems over short times or even in the long-time limit~\cite{FCS2020,CFS2019,P2022,GEF2024,SVH2024,CP2026,CP2026_2}. 
Examples include clock models such as classical pendulum clocks~\cite{P2022}, an electronic clock~\cite{GEF2024}, and a superconducting clock-circuit~\cite{SVH2024}.
While some variants of the TUR have been shown to hold for underdamped systems~\cite{VH2019,LPP2019,LPP2021,D2022}, they depend additionally on dynamical quantities and are thus model-dependent or have generally complicated forms.

The TUR~\eqref{eq.TUR} places a constraint on the fluctuation of a current and the entropy production for nonequilibrium systems. 
This is natural when considering that the current is associated with nonequilibrium processes with dissipation.
Yet, a current is not exclusive to nonequilibrium systems: 
A frictionless pendulum shows a neutrally-stable periodic oscillation sustained 
even without a mechanism for energy injection and dissipation~\cite{Stz2001}.
When representing the pendulum's motion in terms of amplitude and phase, the phase, which increases with time, may thus be regarded as a {\it reversible} current.
In the presence of friction and noise, time can still be measured by the phase of a stochastic pendulum clock, whose oscillation is sustained by an escapement mechanism~\cite{DX2013}.
The TUR~\eqref{eq.TUR} may trivially be violated for such a phase and its fluctuation, particularly when the reversible component, which does not contribute to entropy production, dominates.
This consideration raises a fundamental question: Are there any universal constraints applicable to general stochastic pendulum clocks beyond the conventional TUR?

In this Letter, we show that under several assumptions the long-time uncertainty product of general stochastic pendulum clocks takes a simple form that depends solely on the degree of nonlinearity. 
This relation between the precision and entropy production of stochastic pendulum clocks is universal in underdamped systems in which the TUR~(\ref{eq.TUR}) can be violated.
We verify our theory using several representative models of pendulum clocks, 
thereby demonstrating both its universality and limitations.
The finding of such a universal and simple constraint may open up a new possibility to design elaborate pendulum clocks that are capable of operating efficiently in stochastic environments.

{\sl Model}--.
We adopt a weakly nonlinear oscillator~\cite{Stz2001} as the simplest model for stochastic pendulum clocks:
\begin{align}
\ddot \theta=-\theta+\varepsilon h(\theta,\dot \theta)+\sqrt{2D}\xi,\label{eq.pendulum}
\end{align}
where $\theta$ is the angle of the pendulum, the overdot denotes the time derivative, and $\dot \theta$ denotes the angular velocity.
The function $h(\theta, \dot \theta)$ is defined as
\begin{eqnarray}
h(\theta, \dot \theta)\equiv -\dot \theta+f(\theta, \dot \theta),\label{eq.force}
\end{eqnarray}
where $\varepsilon>0$ denotes a small dimensionless parameter characterizing the degree of nonlinearity. 
Here, the first term on the right-hand side of Eq.~(\ref{eq.force}) represents the frictional force, while the effect of nonlinearity generated, for example, by an escapement mechanism is represented by $f(\theta, \dot \theta)$.
The white Gaussian noise $\xi$, which satisfies $\left<\xi(t)\right>=0$ and $\left<\xi(t)\xi(t')\right>=\delta(t-t')$, is added, where $D\equiv \varepsilon T$ is the diffusion coefficient with $T$ the temperature of a thermal environment.
Note that when $T=0$ the equation of motion~(\ref{eq.pendulum}) reduces to the deterministic one, which is assumed to possess a limit cycle solution that represents self-sustained oscillation~\cite{Stz2001}.

{\sl Main results}--.
For the model described by Eqs.~(\ref{eq.pendulum}) and (\ref{eq.force}) under several additional assumptions, we derive the following long-time uncertainty product as the main result: 
\begin{align}
Q_\infty \equiv \lim_{t\to \infty} \frac{{\rm Var}J_\Phi(t)}{\left<J_\Phi(t)\right>^2} \sigma_{\rm irr}t\approx \frac{\varepsilon^2}{2},\label{eq.TUR_osc}
\end{align}
which takes a universal and simple form irrespective of the form of $f(\theta, \dot \theta)$ and depends solely on $\varepsilon$.

Here, as the time-integrated current $J(t)$, we adopt an accumulated phase $J_\Phi(t)$ in the phase space of the pendulum clock that measures the advancement of time:
The Cartesian coordinates $\bm x=(\theta, \omega)$ with $\omega=\dot \theta$ of the pendulum clock can be written as
$\bm x=(\theta, \omega)=(R\cos \Phi, R\sin \Phi)$
in polar coordinates with amplitude $R(\bm x)$ and phase $\Phi(\bm x)$ defined as
\begin{align}
&R(\bm x)\equiv \sqrt{\theta^2+{\omega}^2},\label{eq.def_R}\\
&\Phi(\bm x)\equiv \arctan \left(\omega/\theta \right).\label{eq.def_Phi}
\end{align}
The accumulated phase is thus defined as
\begin{align}
J_\Phi(t)\equiv \int_0^t ds \bm \Lambda(\bm x) \circ \dot{\bm x}=\int_0^t ds \nabla \Phi \circ \dot{\bm x},\label{eq.J}
\end{align}
where $\circ$ denotes the Stratonovich product and $\bm \Lambda(\bm x)$ is a projection function, 
where we adopt $\bm \Lambda(\bm x)=\nabla \Phi(\bm x)$ in terms of the gradient of the phase~(\ref{eq.def_Phi}):
\begin{eqnarray}
\nabla \Phi=\left(-\frac{\omega}{\theta^2+{\omega}^2}, \frac{\theta}{\theta^2+{\omega}^2}\right)^{\mathrm T}.
\end{eqnarray}
The time-integrated current~\eqref{eq.J} involves the product of the projection function and the change rate of a system's state variables $\bm x$ including the velocity degree of freedom.
Although such a definition of a current may not be conventional (but see~\cite{VH2019}), it is natural to take the accumulated phase as a time-integrated current here.

By taking the ensemble average of Eq.~(\ref{eq.J}), we can show 
\begin{eqnarray}
\left<J_\Phi(t)\right>&&=\left(\int d\bm x(\nabla \Phi)^{\rm T} \bm j(\bm x)\right)t=\Omega t,\label{eq.J_avr}
\end{eqnarray}
where
\begin{eqnarray}
\Omega \equiv -1+\left(\int d\bm x(\nabla \Phi)^{\rm T}\bm j_{\rm irr}(\bm x)\right).\label{eq.def_Omega}
\end{eqnarray}
Here, $\bm j(\bm x)=(j_\theta(\bm x), j_\omega(\bm x))^{\mathrm T}$ is the stationary probability current in the Fokker-Planck equation for the probability distribution of $\bm x$ corresponding to Eq.~\eqref{eq.pendulum},
which can be decomposed into reversible and irreversible parts~\cite{SF2012}:
\begin{eqnarray}
\bm j(\bm x)=\bm j_{\rm rev}(\bm x)+\bm j_{\rm irr}(\bm x),\label{eq.J_decomp}
\end{eqnarray}
where only the irreversible part $\bm j_{\rm irr}(\bm x)$ contributes to the entropy production as given below.
In Eq.~(\ref{eq.def_Omega}), we used $\int d\bm x(\nabla \Phi)^{\rm T}\bm j_{\rm rev}(\bm x)=-1$.
See Sec.~I of Supplemental Material (SM)~\cite{SM} for the derivation of Eq.~(\ref{eq.J_avr}).

The entropy production rate $\sigma_{\rm irr}$ in Eq.~(\ref{eq.TUR_osc}) is given as follows.
For systems with velocity-dependent forces as the present case, the entropy production rate for steady states reads~\cite{KQ2004,LPP2019,HSE2020}:
\begin{align}
\sigma_{\rm irr}=\frac{1}{D} \int d{\bm x} \frac{\bm j_{\rm irr}^2(\bm x)}{p({\bm x})}=-\frac{\mathcal J_Q}{T}-\sigma_{\rm pump}\ge 0,\label{eq.sigma_def}
\end{align}
where $p(\bm x)$ denotes the stationary probability distribution.
It arises solely from the irreversible probability current and consists of the entropy transfer $-\mathcal J_Q/T$ from a system to a thermal environment and an entropy pumping effect $\sigma_{\rm pump}$. 
The entropy transfer is given by
\begin{align}
-\frac{\mathcal J_Q}{T}=\frac{2\varepsilon}{T}\left(\frac{T_{\rm eff}}{2}-\frac{T}{2}\right),\label{eq.entropy_transfer}
\end{align}
where $\mathcal J_Q$ denotes the heat from the thermal environment to the system and we defined the effective temperature of the pendulum clock by $T_{\rm eff}\equiv \left<R^2\right>/2$. 
The entropy pumping is given by
\begin{align}
\sigma_{\rm pump}\equiv \varepsilon \int  d{\bm x} \frac{\partial f({\bm x})}{\partial \omega} p({\bm x}),
\end{align}
which arises only for velocity-dependent forces.

\begin{table*}[t!]
\caption{Examples of stochastic pendulum clocks.}
\label{table_escapement}
\begin{ruledtabular}
\begin{tabular}{lllr}
Model & Force $f(\theta, \dot \theta)$ & Phase average $\bar f_R(R)$ & Radius of limit cycle $R^*$ \\
\colrule 
Grasshopper escapement~\cite{H2014,Z2021,Z2022} & $-{\rm sgn}(\theta-\theta_r {\rm sgn}(\dot \theta))$ & $2\theta_r/\pi R$ & $\sqrt{4\theta_r/\pi}$\\
Graham escapement~\cite{H2014} & $-{\rm sgn}(\sin \psi_0 \theta-\cos \psi_0 \dot \theta)$ & $2\cos \psi_0/\pi$ & $4\cos \psi_0/\pi$\\
van der Pol oscillator~\cite{Stz2001} & $(2-\theta^2)\dot \theta$ & $R^3/8$ & 2
\end{tabular}
\end{ruledtabular}
\end{table*}

By substituting Eq.~(\ref{eq.J_avr}) into Eq.~(\ref{eq.TUR_osc}), we have
\begin{align}
Q_\infty=\frac{2D_\Phi}{\Omega^2}\sigma_{\rm irr},\label{eq.TUR_osc_2}
\end{align}
where we introduced the phase diffusion coefficient defined by
$D_{\Phi} \equiv \lim_{t\to \infty} {\rm Var}J_\Phi(t)/2t$
as a measure of the precision of the pendulum clock.

The dynamics of the amplitude $R$ and phase $\Phi$ in Eqs.~(\ref{eq.def_R}) and (\ref{eq.def_Phi}) after phase averaging under the assumption of time-scale separation ($\varepsilon \ll 1$) reads~\cite{M2020,ZY1987} (see Sec.~II of SM~\cite{SM} for the derivation)
\begin{align}
&\dot R=-\frac{\varepsilon R}{2}+\varepsilon \bar f_R(R)+\frac{D}{2R}+\sqrt{2\left(\frac{D}{2}\right)}\xi_R,\label{eq.amplitude}\\
&\dot \Phi=-1+\varepsilon \bar f_\Phi(R)+\sqrt{2\left(\frac{D}{2R^2}\right)} \cdot \xi_\Phi ,\label{eq.phase}
\end{align}
where $\bar f_R(R)\equiv (1/2\pi)\int_0^{2\pi}\sin \Phi f(R\cos \Phi, R\sin \Phi)d\Phi$ and $\bar f_\Phi(R)\equiv (1/2\pi)\int_0^{2\pi}\cos \Phi f(R\cos \Phi, R\sin \Phi)/Rd\Phi$ denote the phase-averaged forces, $\xi_R$ and $\xi_\Phi$ denote the independent white Gaussian noises, and the stochastic product $\cdot$ denotes the It\^o product.

In addition to the time-scale separation ($\varepsilon \ll 1$), by further assuming that the radius of the stochastic limit cycle may be replaced with its deterministic value $R^*$ in Eq.~(\ref{eq.phase}), we find that the phase diffusion coefficient $D_\Phi$ is approximated by
\begin{align}
D_\Phi \simeq \frac{D}{2{R^*}^2}=\frac{\varepsilon T}{2{R^*}^2}.\label{eq.D_Temp}
\end{align}
Here, $R^*$ is determined by the solution of the following equation: 
\begin{eqnarray}
-\frac{R^*}{2}+\bar f_R(R^*)=0\label{eq.R_sol}
\end{eqnarray}
as the fixed point of Eq.~(\ref{eq.amplitude}) with $T=0$. 
Equation (\ref{eq.D_Temp}) shows that the phase diffusion coefficient is inversely proportional to the square of the radius of the limit cycle,
which is consistent with the result for a pendulum clock model in~\cite{M2020}.
The validity of the above replacement of the stochastic variable $R$ with its deterministic value $R^*$ may be quantified by the coefficient of variation (CV) of $R$:
\begin{eqnarray}
{\rm CV}\equiv \frac{\sqrt{{\rm Var} R}}{\left<R\right>}.\label{eq.def_CV}
\end{eqnarray}
The smaller (larger) CV is, the more deterministic (stochastic) $R$ is. Practically, ${\rm CV}\ll 1$ may be a convenient measure.
One promising design principle is to make $R^*$ sufficiently large.

By further assuming that the system's temperature $T_{\rm eff}$ is sufficiently larger than the environmental one $T$ as $T_{\rm eff}\approx {R^*}^2/2 \gg T$, we can also approximate $\sigma_{\rm irr}$ in Eq.~(\ref{eq.sigma_def}) as (see Sec.~III of SM~\cite{SM} for the derivation)
\begin{eqnarray}
\sigma_{\rm irr}\approx \frac{\varepsilon {R^*}^2}{2T}\approx \frac{\varepsilon^2}{4D_\Phi },\label{eq.sigma}
\end{eqnarray}
where we used Eq.~(\ref{eq.D_Temp}) in the second equality.
Equation (\ref{eq.sigma}) suggests that a smaller (larger) phase diffusion comes at larger (smaller) dissipation, showing their reciprocal relationship.

Using the reciprocal relation Eq.~(\ref{eq.sigma}) between $\sigma_{\rm irr}$ and $D_\Phi$ in Eq.~(\ref{eq.TUR_osc_2}), and $\Omega=-1+O(\varepsilon)$ (see Sec.~IV of SM~\cite{SM} for the derivation), we derive our main result Eq.~(\ref{eq.TUR_osc}).
As $\varepsilon$ can be arbitrarily small, pendulum clocks can trivially violate the TUR (\ref{eq.TUR}):
The usual constraint between the current and entropy production is lost here because the reversible current mainly contributes to the current.

{\sl Demonstrations}--.
\begin{figure*}[t!]
\centering
\includegraphics[scale=0.54]{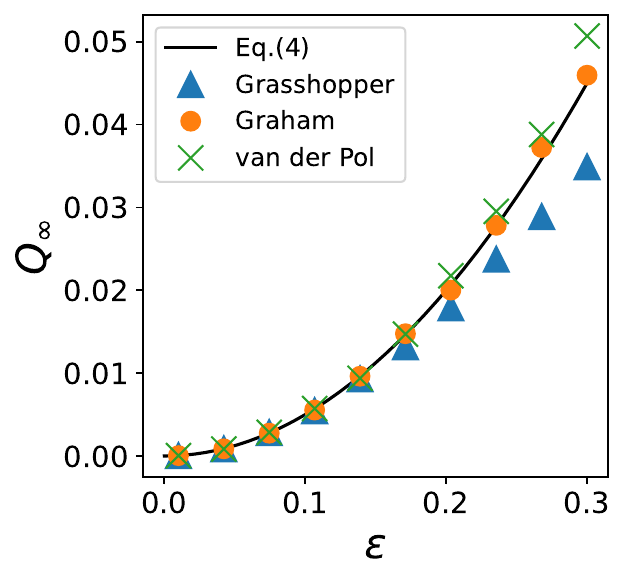}
\hfill
\includegraphics[scale=0.54]{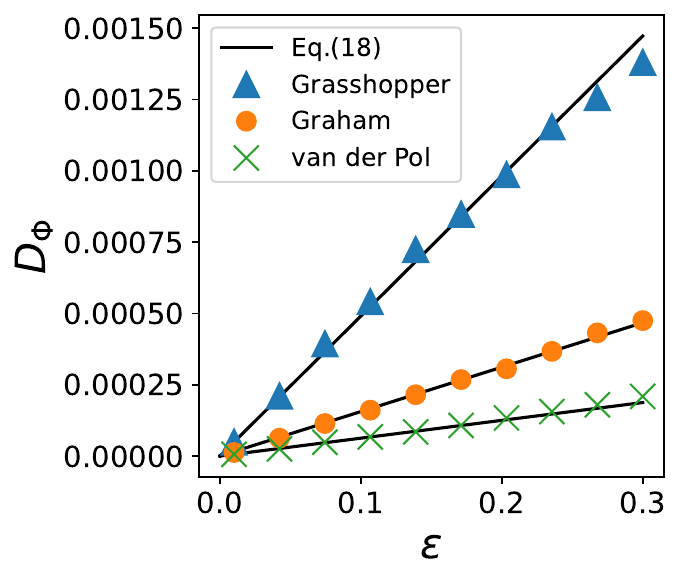}
\hfill
\includegraphics[scale=0.54]{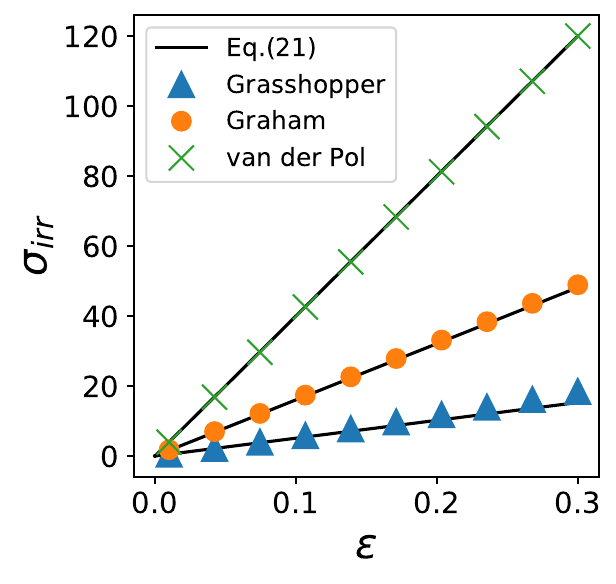}
\caption{Comparison of the theoretical predictions for (a) the long-time uncertainty product $Q_\infty$ in Eq.~(\ref{eq.TUR_osc}), (b) the phase diffusion coefficient $D_\Phi$ in Eq.~(\ref{eq.D_Temp}), and (c) the entropy production rate $\sigma_{\rm irr}$ in Eq.~(\ref{eq.sigma}), with the corresponding numerical results for the three models in Table~\ref{table_escapement}. 
In the numerical simulation, we replaced the sign function ${\rm sgn}(x)$ with $\tanh(nx)$ by noting $\lim_{n\to \infty} \tanh(nx)={\rm sgn}(x)$. We used the Euler-Maruyama method with time step $dt=0.001$ for numerical integration and $4000$ samples for the ensemble averages. The parameters are chosen as $T=0.005$, $\theta_r=0.4$, $\psi_0=0.1$, and $n=20$.}
\label{fig}
\end{figure*}
\begin{figure}
\centering
\includegraphics[scale=0.75]{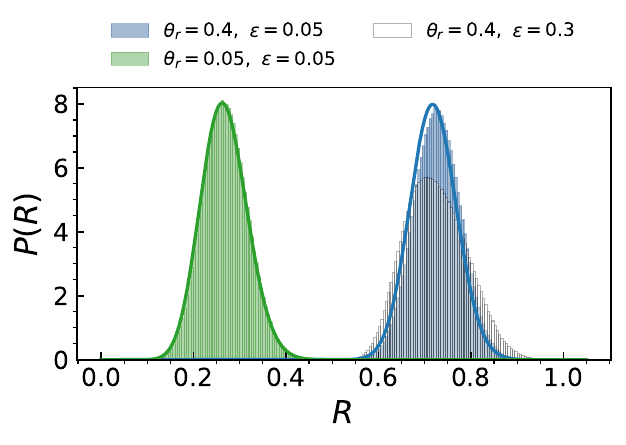}
\caption{Stationary probability distribution $P(R)$ for the Grasshopper-escapement model with different $\varepsilon$ and $\theta_r$, 
compared with the analytical distribution given by Eq.~(\ref{eq.P_st}) under the time-scale separation ($\varepsilon \ll 1$) (solid curves). 
The numerical method and all other parameters are the same as in Fig.~\ref{fig}.}
\label{fig_distribution}
\end{figure}
\begin{figure}
\centering
\includegraphics[scale=0.55]{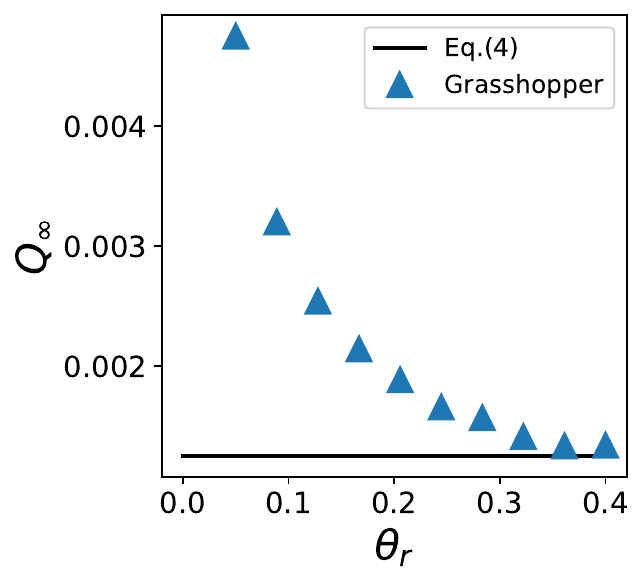}
\caption{The long-time uncertainty product $Q_\infty$ of the Grasshopper-escapement model as a function of $\theta_r$ for $\varepsilon=0.05$, compared with the theoretical line $Q_\infty \approx \varepsilon^2/2$ (Eq.~(\ref{eq.TUR_osc})). The numerical method, sample size, and all other parameters are the same as in Fig.~\ref{fig}.}
\label{fig.small_theta}
\end{figure}
\begin{figure}
\centering
\includegraphics[scale=0.55]{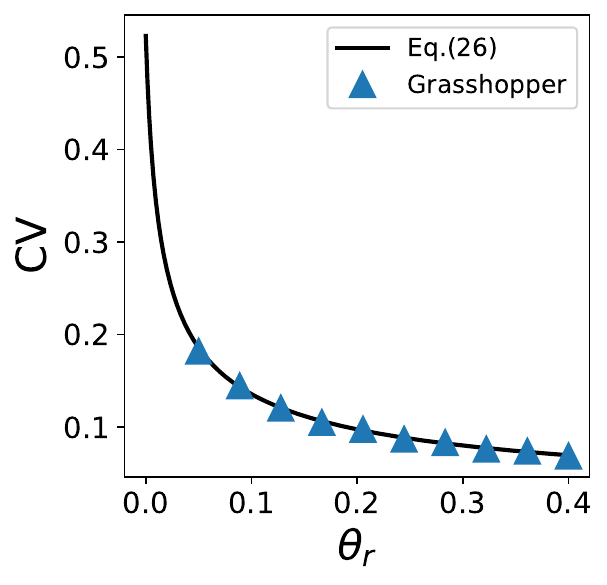}
\caption{The CV in Eq.~(\ref{eq.def_CV}) of the Grasshopper-escapement model as a function of $\theta_r$ for $\varepsilon=0.05$, compared with the theoretical curve given by Eq~(\ref{eq.CV_theory}). The numerical method, sample size, and all other parameters are the same as in Fig.~\ref{fig}.}
\label{fig.small_theta_CV}
\end{figure}
We adopt three stochastic pendulum clock models to demonstrate the main result Eq.~(\ref{eq.TUR_osc}).
Each model is equipped with a mechanism that allows self-sustained oscillation in its force term.

The first example adopts the Grasshopper escapement described by~\cite{H2014,Z2021,Z2022}:
\begin{align}
f(\theta, \dot \theta)=-{\rm sgn}(\theta-\theta_r {\rm sgn}(\dot \theta)),\label{eq.grass}
\end{align}
where $\mathrm{sgn}(\cdot)$ denotes the sign function and $\theta_r>0$ denotes a critical angle. 
The constant force is applied in the same direction as $\dot{\theta}$, thereby assisting the pendulum's motion, but its direction is reversed when the pendulum exceeds the critical angle $\theta_r$, i.e., when $\theta > \theta_r$ for $\dot{\theta} > 0$ and $\theta < -\theta_r$ for $\dot{\theta} < 0$.

The second example adopts the Graham escapement described by~\cite{H2014}:
\begin{align}
f(\theta, \dot \theta)=-{\rm sgn}(\sin \psi_0 \theta-\cos \psi_0 \dot \theta),\label{eq.graham}
\end{align}
where $0 \le  \psi_0 <\pi/2$. The constant force is applied, whose sign changes every half cycle upon crossing the line $\dot \theta=(\tan \psi_0)\theta$ in the phase plane. In particular, when $\psi_0=0$, the force is applied exactly in the same direction as $\dot \theta$.

As the third example, we adopt the following forcing:
\begin{eqnarray}
f(\theta, \dot \theta)=(2-\theta^2)\dot \theta.\label{eq.vanderpol}
\end{eqnarray}
With this $f(\theta, \dot \theta)$, we recover a nonlinear state-dependent frictional force $h(\theta, \dot \theta)=(1-\theta^2)\dot \theta$ of the van der Pol oscillator, which is a representative model of the weakly nonlinear oscillator~\cite{Stz2001}.
Depending on the magnitude of $\theta$, the nonlinear friction $1-\theta^2$ changes its sign, thereby sustaining the oscillation by the injection and dissipation of the energy.

In Table~\ref{table_escapement}, we summarize the phase-average $\bar f_R(R)$ and the radius of the limit cycle $R^*$ as the solution of Eq.~(\ref{eq.R_sol}) for each model.

In Fig.~\ref{fig}, we show the numerical results of Eqs.~(\ref{eq.TUR_osc}), (\ref{eq.D_Temp}), and (\ref{eq.sigma}) for each model. 
We find that the long-time uncertainty product for each model is in good agreement with the formula in Eq.~(\ref{eq.TUR_osc}) for sufficiently small $\varepsilon$.
Meanwhile, the phase diffusion coefficient and entropy production rate are model-dependent through the limit-cycle radius as predicted by Eqs.~(\ref{eq.D_Temp}) and (\ref{eq.sigma}).

As shown in Fig.~\ref{fig}, the long-time uncertainty product shows discrepancies between the numerical results and the formula in Eq.~(\ref{eq.TUR_osc}) for large $\varepsilon$.
This is because the assumption of the time-scale separation (Eqs.~(\ref{eq.amplitude}) and (\ref{eq.phase})) is no longer valid for large $\varepsilon$.
To confirm this prediction, we show in Fig.~\ref{fig_distribution} the stationary probability distribution of $R$ for the Grasshopper escapement and compare it with the analytical form obtained from Eq.~(\ref{eq.amplitude}):
\begin{eqnarray}
P(R) = \frac{2}{(2T)^{\frac{2\theta_r}{\pi T}+1}\,\Gamma\!\left(\frac{2\theta_r}{\pi T}+1\right)}\; R^{\frac{4\theta_r}{\pi T}+1}\,\exp\!\left(-\frac{R^2}{2T}\right),\label{eq.P_st}\nonumber\\
\end{eqnarray}
where $\Gamma(z)\equiv \int_0^\infty e^{-s}s^{z-1} ds$ denotes the Gamma function.
Note that Eq.~(\ref{eq.P_st}) is independent of $\varepsilon$. 
By comparing the two distributions for $\varepsilon=0.05$ and $\varepsilon=0.3$, we find that the numerical result and Eq.~(\ref{eq.P_st}) agree well for $\varepsilon=0.05$, while they do not for $\varepsilon=0.3$ as expected, indicating the violation of the assumption of the time-scale separation.

Even if the assumption of the time-scale separation is valid, the assumption of the replacement of the stochastic radius with its deterministic value is also necessary for the formula Eq.~(\ref{eq.TUR_osc}) to hold.
In fact, when $\theta_r$ in Eq.~(\ref{eq.grass}) is small, while its stationary probability distribution is still well approximated by Eq.~(\ref{eq.P_st}) (Fig.~\ref{fig_distribution}), its CV (\ref{eq.def_CV}) may increase as $R^*$ decreases as in Table~\ref{table_escapement}, leading to the violation of Eq.~(\ref{eq.TUR_osc}).
To confirm this prediction, we show in Fig.~\ref{fig.small_theta} the numerical result of $Q_\infty$ for small $\varepsilon$ as a function of $\theta_r$ that controls the radius of the limit cycle. 
We find that as $\theta_r$ decreases, the deviation from the theory increases as expected, justifying the necessity of the present assumption.

As supportive evidence, the CV as a function of $\theta_r$ is also shown in Fig.~\ref{fig.small_theta_CV}. 
The numerical results are compared with the theoretical curve:
\begin{eqnarray}
{\rm CV}=\sqrt{\frac{\alpha}{2}\Biggl[\frac{\Gamma(\frac{\alpha}{2})}{\Gamma(\frac{\alpha+1}{2})}\Biggr]^2-1}, \ \alpha=\frac{4\theta_r}{\pi T}+2,\label{eq.CV_theory}
\end{eqnarray}
which is calculated by using Eq.~(\ref{eq.P_st}). 
As expected, the CV decreases as $\theta_r$ increases.
Note that the decrease in the CV is relatively slow because of the asymptotic inverse-square-root scaling ${\rm CV}\simeq \sqrt{1/2\alpha}$ for $\theta_r/T \gg 1$.

{\sl Concluding perspective}--.
We established a simple law governing stochastic pendulum clocks, an important class of underdamped systems for which the TUR~(\ref{eq.TUR}) can be violated.
We showed that the long-time uncertainty product between the precision and entropy production of stochastic pendulum clocks takes the universal form given by Eq.~(\ref{eq.TUR_osc}).
We demonstrated it using several representative stochastic pendulum clock models, and its limitations was also addressed.
Given its fundamental nature and broad applicability, our result is expected to contribute to the thermodynamics of clocks.

We end this Letter by suggesting interesting directions for further research.
Detailed characterization of the uncertainty product in the noisy regime where the time-scale separation holds but the fluctuation of the phase diffusion coefficient is significant is important.
In this regime, as the amplitude fluctuates randomly independently of the phase, the phase diffusion coefficient itself may behave as a random variable (Eq.~(\ref{eq.phase})), which may call for the concept of diffusing diffusivity~\cite{CS2014,CSMS2017}.
Although we adopted the simplest one-dimensional model (\ref{eq.pendulum}), more complex higher-dimensional models that incorporate the degrees of freedom associated with both the pendulum and the escapement are also commonly used~\cite{P2022,GEF2024,SVH2024}. It is an interesting task to reduce a high-dimensional model to a one-dimensional one by adiabatically eliminating the degrees of freedom of the escapement as a fast variable, thereby a momentum-dependent force term may appear as in Eq.~(\ref{eq.pendulum}).
Investigation of coupled pendulum clocks or more generally coupled oscillatory systems has long been an active area of research~\cite{K1984,PRK2001}.
Elucidating how collective effects such as synchronization affect the precision and energetics of stochastic pendulum clocks is a promising direction for future research~\cite{IKS2016,LHJ2018,CEP2026}.

\acknowledgments
This work was supported by JSPS KAKENHI Grant Number 25K07163.
The author used ChatGPT (OpenAI, GPT-5.5 and GPT-5.6 Sol) to assist with editing the manuscript and debugging
numerical codes used for the numerical calculations.

\clearpage
\onecolumngrid
\setcounter{equation}{0} 

\begin{center}
\textbf{\large Supplemental Material for ``A Universal Thermodynamic Law Governing Stochastic Pendulum Clocks"}
\end{center}

\begin{center}
Yuki Izumida\\
{\it Department of Complexity Science and Engineering, Graduate School of Frontier Sciences, The University of Tokyo, Kashiwa 277-8561, Japan}
\end{center}
\vspace{5pt}

In this Supplemental Material, we provide derivations of some of the equations in the main text.
\section{I. Derivation of Eq.~(9)}
The Fokker-Planck equation corresponding to Eq.~(2) in the main text reads~\cite{US2025_S}
\begin{eqnarray}
\frac{\partial p(\bm x, t)}{\partial t}=-\nabla \cdot \bm j(\bm x, t),
\end{eqnarray}
where $p(\bm x, t)$ denotes the probability distribution of $\bm x=(\theta, \omega)$ at time $t$ and $\bm j(\bm x, t)=(j_\theta(\bm x, t), j_\omega(\bm x, t))^{\mathrm T}$ denotes the probability current:
\begin{eqnarray}
&&j_\theta(\bm x, t)=\omega p(\bm x, t),\\
&&j_\omega(\bm x, t)=\left(-\theta+\varepsilon f(\bm x)-\varepsilon \omega-D\frac{\partial}{\partial \omega}\right)p(\bm x, t).
\end{eqnarray}
We consider the stationary state satisfying $\partial_t p(\bm x, t)=0$, and denote the stationary probability distribution by $p(\bm x)$ and the stationary probability current by $\bm j(\bm x)$, respectively.
The reversible and irreversible parts $\bm j_{\rm rev}(\bm x)=(j_{{\rm rev}, \theta}(\bm x), j_{{\rm rev}, \omega}(\bm x))^{\mathrm T}$ and $\bm j_{\rm irr}(\bm x)=(j_{{\rm irr}, \theta}(\bm x), j_{{\rm irr}, \omega}(\bm x))^{\mathrm T}$ of the stationary probability current $\bm j(\bm x)=\bm j_{\rm rev}(\bm x)+\bm j_{\rm irr}(\bm x)$ are given by
\begin{align}
&j_{{\rm rev}, \theta}(\bm x)=\omega p(\bm x),\label{eq.Jrev_theta_S}\\
&j_{{\rm rev}, \omega}(\bm x)=-\theta p(\bm x),\label{eq.Jrev_omega_S}\\
&j_{{\rm irr}, \theta}(\bm x)=0,\label{eq.Jirr_theta_S}\\
&j_{{\rm irr}, \omega}(\bm x)=\left(\varepsilon f(\bm x)-\varepsilon \omega-D\frac{\partial}{\partial \omega}\right)p(\bm x).\label{eq.Jirr_omega_S}
\end{align}
By taking the ensemble average of $J_\Phi(t)$ in Eq.~(7) in the main text, we have
\begin{align}
\left<J_\Phi(t)\right>&=\left(\int d\bm x(\nabla \Phi)^{\rm T} \bm j(\bm x)\right)t=\left(\int d\bm x(\nabla \Phi)^{\rm T}(\bm j_{\rm rev}(\bm x)+\bm j_{\rm irr}(\bm x))\right)t.
\end{align}
By using Eq.~(8) in the main text and Eqs.~(\ref{eq.Jrev_theta_S}) and (\ref{eq.Jrev_omega_S}), we obtain $\int d\bm x(\nabla \Phi)^{\rm T}\bm j_{\rm rev}(\bm x)=-1$, and thus derived Eq.~(9) in the main text.

\section{II. Derivation of Eqs.~(16) and (17)}
By using the equation of motion Eq.~(2) and the definition of the amplitude $R$ and phase $\Phi$ in Eqs.~(5) and (6) in the main text, 
we can obtain their dynamics as
\begin{align}
&\dot R=-\varepsilon R\sin^2 \Phi+\varepsilon \sin \Phi f(R\cos \Phi, R\sin \Phi)+\sqrt{2D}\sin \Phi \circ \xi,\label{eq.R_S}\\
&\dot \Phi=-1-\varepsilon \cos \Phi \sin \Phi+\frac{\varepsilon \cos \Phi f(R\cos \Phi, R\sin \Phi)}{R}+\frac{\sqrt{2D}\cos \Phi}{R}\circ \xi,\label{eq.Phi_S}
\end{align}
respectively.
For convenience, we transform Eqs.~(\ref{eq.R_S}) and (\ref{eq.Phi_S}) into the It\^o type:
\begin{align}
&\dot R=-\varepsilon R\sin^2 \Phi+\varepsilon \sin \Phi f(R\cos \Phi, R\sin \Phi)+\frac{D\cos^2 \Phi}{R}+\sqrt{2D}\sin \Phi \cdot \xi,\label{eq.R_S2}\\
&\dot \Phi=-1-\varepsilon \cos \Phi \sin \Phi+\frac{\varepsilon \cos \Phi f(R\cos \Phi, R\sin \Phi)}{R}-2D\frac{\cos \Phi \sin \Phi}{R^2}+\frac{\sqrt{2D}\cos \Phi}{R}\cdot \xi,\label{eq.Phi_S2}
\end{align}
where the dot product refers to the It\^o product.
Note that because $\xi$ appears in both the dynamics of $R$ and $\Phi$, the diffusion matrix corresponding to the noise terms in Eqs.~(\ref{eq.R_S2}) and (\ref{eq.Phi_S2}) includes the non-diagonal elements:
\begin{align}
\left(
\begin{array}{cc}
D_{RR} & D_{R\Phi} \\
D_{\Phi R} & D_{\Phi \Phi}
\end{array}
\right)
=2D
\left(
\begin{array}{cc}
\sin^2 \Phi & \sin \Phi \cos \Phi/R \\
\sin \Phi \cos \Phi/R & \cos^2 \Phi/R^2
\end{array}
\right).\label{SD}
\end{align}
We can perform a phase averaging with respect to the fast variable $\Phi$ as the time-scales between $\Phi$ and $R$ are sufficiently separated for $\varepsilon \ll 1$.
By phase-averaging Eqs.~(\ref{eq.R_S2}) and (\ref{eq.Phi_S2}), we derive Eqs.~(16) and (17) in the main text:
\begin{align}
&\dot R=-\frac{\varepsilon R}{2}+\frac{\varepsilon}{2\pi}\int_0^{2\pi}\sin \Phi f(R\cos \Phi, R\sin \Phi)d\Phi+\frac{D}{2R}+\sqrt{2\left(\frac{D}{2}\right)} \cdot \xi_R,\label{eq.R_S3}\\
&\dot \Phi=-1+\frac{\varepsilon}{2\pi}\int_0^{2\pi}\frac{\cos \Phi f(R\cos \Phi, R\sin \Phi)}{R}d\Phi+\sqrt{2\left(\frac{D}{2R^2}\right)}\cdot \xi_\Phi,\label{eq.Phi_S3}
\end{align}
where the non-diagonal elements of the diffusion matrix (\ref{SD}) vanish upon the phase average:
\begin{align}
\left(
\begin{array}{cc}
\bar D_{RR} & \bar D_{R\Phi} \\
\bar D_{\Phi R} & \bar D_{\Phi \Phi}
\end{array}
\right)
=
\left(
\begin{array}{cc}
\frac{1}{2\pi}\int_0^{2\pi}D_{RR}d\Phi & \frac{1}{2\pi}\int_0^{2\pi}D_{R\Phi}d\Phi  \\
\frac{1}{2\pi}\int_0^{2\pi}D_{\Phi R}d\Phi & \frac{1}{2\pi}\int_0^{2\pi}D_{\Phi \Phi}d\Phi 
\end{array}
\right)
=2D
\left(
\begin{array}{cc}
1/2  & 0 \\
0 & 1/2R^2
\end{array}
\right).\label{SD_avr}
\end{align}

\section{III. Derivation of Eq.~(21)}
We derive Eq.~(21) in the main text. 
We show that the main contribution to $\sigma_{\rm irr}$ in Eq.~(12) is given by the entropy transfer $-\mathcal J_Q/T$ in Eq.~(13) and $\sigma_{\rm pump}$ in Eq.~(14) can be neglected.

First, we show that $-\mathcal J_Q/T=O(\varepsilon T^{-1})$ when $T_{\rm eff} \simeq {R^*}^2/2 \gg T$ is assumed. From the definition in Eq.~(14), we have
\begin{align}
-\frac{\mathcal J_Q}{T}&=\frac{2\varepsilon}{T}\left(\frac{T_{\rm eff}}{2}-\frac{T}{2}\right)\approx \frac{2\varepsilon}{T}\left(\frac{{R^*}^2}{4}-\frac{T}{2}\right) \simeq \frac{\varepsilon {R^*}^2}{2T}=O(\varepsilon T^{-1}), \label{eq.entropy_transfer_S}
\end{align}
where we used $T_{\rm eff} \simeq {R^*}^2/2$ in the second equality and used $ {R^*}^2/2 \gg T$ in the third equality.

Next, we show $\sigma_{\rm pump}=O(\varepsilon)$ and can thus be neglected compared to $-\mathcal J_Q/T=O(\varepsilon T^{-1})$. From the definition Eq.~(13), we have
\begin{align}
\sigma_{\rm pump}=\varepsilon \int  d{\bm x} \frac{\partial f({\bm x})}{\partial \omega} p({\bm x})
&=\varepsilon \int  dRd\Phi \ \frac{\partial f({\bm x})}{\partial \omega}\biggl |_{{\bm x}={\bm x}(R, \Phi)} P(R, \Phi)\nonumber\\
&\approx \varepsilon \int  dRd\Phi \ \frac{\partial f({\bm x})}{\partial \omega}\biggl |_{{\bm x}={\bm x}(R, \Phi)}\frac{1}{2\pi} \delta(R-R^*)\nonumber\\
&=\frac{\varepsilon}{2\pi} \int_0^{2\pi} d\Phi \ \frac{\partial f({\bm x})}{\partial \omega}\biggl |_{{\bm x}={\bm x}(R^*, \Phi)}\nonumber\\
&=O(\varepsilon).
\end{align}
Here, we have used $p(\bm x)d\bm x=p(\theta, \omega)d\theta d\omega=dRd\Phi P(R, \Phi)$ with $P(R, \Phi)$ the stationary probability distribution of $R$ and $\Phi$ in the second equality, and $P(R, \Phi)\approx P(R)P(\Phi)=\delta(R-R^*)/2\pi$ in the third equality.
This completes the derivation of Eq.~(21).

\section{IV. Approximation to Eq.~(10)}
By substituting Eqs.~(\ref{eq.Jirr_theta_S}) and (\ref{eq.Jirr_omega_S}) into Eq.~(10) in the main text, we have
\begin{align}
\Omega&=-1+\int d\bm x (\nabla \Phi)^{\rm T}\bm j_{\rm irr}(\bm x)\nonumber\\
&=-1+\int_{-\infty}^\infty \int_{-\infty}^\infty \frac{\theta}{\theta^2+\omega^2} j_{{\rm irr}, \omega}(\bm x)d\theta d\omega \nonumber \\
&=-1+\int_{-\infty}^\infty \int_{-\infty}^\infty \biggl [\frac{\varepsilon \theta f(\theta, \omega)}{\theta^2+\omega^2}p(\theta, \omega)-\frac{\varepsilon \theta \omega}{\theta^2+\omega^2}p(\theta, \omega)-D\frac{\theta}{\theta^2+\omega^2}\frac{\partial p(\theta, \omega)}{\partial \omega} \biggr ]d\theta d\omega \nonumber\\
&=-1+\int_{-\infty}^\infty \int_{-\infty}^\infty \frac{\varepsilon \theta f}{\theta^2+\omega^2}p(\theta, \omega)d\theta d\omega-\int_{-\infty}^\infty \int_{-\infty}^\infty \frac{\varepsilon \theta \omega}{\theta^2+\omega^2}p(\theta, \omega)d\theta d\omega-\int_{-\infty}^\infty \underbrace{\biggl [D\frac{\theta}{\theta^2+\omega^2}p(\theta, \omega) \biggr ]_{-\infty}^\infty}_{=0} d\theta \nonumber\\
&+D\int_{-\infty}^\infty \int_{-\infty}^\infty \frac{2\theta \omega}{(\theta^2+\omega^2)^2}p(\theta, \omega)d\theta d\omega \nonumber\\
&=-1+\int_0^\infty \int_0^{2\pi}\frac{\varepsilon \cos \Phi f(R\cos \Phi, R\sin \Phi)}{R} P(R, \Phi) dRd\Phi-\int_0^\infty \int_0^{2\pi} \varepsilon \cos\Phi \sin \Phi P(R, \Phi)dRd\Phi\nonumber\\
&+D \int_0^\infty \int_0^{2\pi} \frac{2\cos \Phi \sin \Phi}{R^2} P(R, \Phi)dRd\Phi\nonumber\\
&\approx -1+\frac{\varepsilon}{2\pi}\int_0^{2\pi} \frac{\cos \Phi f(R^*\cos \Phi, R^*\sin \Phi)}{R^*}d\Phi\nonumber\\
&=-1+\varepsilon \bar f_\Phi(R^*)\nonumber\\
&=-1+O(\varepsilon).
\end{align}
In the third to last equality, we used $P(R, \Phi)\approx P(R)P(\Phi)=\delta(R-R^*)/2\pi$ and $\int_0^{2\pi}\cos \Phi \sin \Phi d\Phi=0$.


\begin{references}
\bibitem{M2020} G. J. Milburn, The thermodynamics of clocks, Contemp. Phys. {\bf 61}, 69 (2020).
\bibitem{EMSWB2017} P. Erker,  M. T. Mitchison, R. Silva, M. P. Woods, N. Brunner, and M. Huber, Autonomous Quantum Clocks: Does Thermodynamics Limit Our Ability to Measure Time?, Phys. Rev. X {\bf 7}, 031022 (2017).
\bibitem{PGELBHA2021} A. N. Pearson, Y. Guryanova, P. Erker, E. A. Laird, G. A. D. Briggs, M. Huber, and N. Ares, Measuring the Thermodynamic Cost of Timekeeping, Phys. Rev. X {\bf 99}, 021029 (2021).
\bibitem{MMSAEG2025} F. Meier, Y. Minoguchi, S. Sundelin, T. J. G. Apollaro, P. Erker, and S. Gasparinetti, Precision is not limited by the second law of thermodynamics, Nat. Phys. {\bf 21}, 1147 (2025).
\bibitem{US2025} U. Seifert, {\it Stochastic Thermodynamics} (Cambridge University Press, Cambridge, 2025).
\bibitem{BS2015} A. C. Barato and U. Seifert, Thermodynamic uncertainty relation for biomolecular processes, Phys. Rev. Lett. {\bf 114}, 158101 (2015).
\bibitem{GHPE2016} T. Gingrich, J. M. Horowitz, N. Perunov, and J. L. England, Dissipation bounds all steady-state current fluctuations, Phys. Rev. Lett. {\bf 116}, 120601 (2016).
\bibitem{PRS2017} P. Pietzonka, F. Ritort, and U. Seifert, Finite-time generalization of the thermodynamic uncertainty relation, Phys. Rev. E {\bf 96}, 012101 (2017).
\bibitem{HG2020} J. M. Horowitz and T. R. Gingrich, Thermodynamic uncertainty relations constrain non-equilibrium fluctuations, Nat. Phys. {\bf 16}, 15 (2020).
\bibitem{DS2018} A. Dechant and S.-i. Sasa, Current fluctuations and transport efficiency for general Langevin systems, J. Stat. Mech. {\bf 2018}, 063209 (2018).
\bibitem{HV2019} Y. Hasegawa and T. Van Vu, Uncertainty relations in stochastic processes: An information inequality approach, Phys. Rev. E {\bf 99}, 062126 (2019).
\bibitem{HV2019_2} Y. Hasegawa and T. Van Vu, Fluctuation Theorem Uncertainty Relation, Phys. Rev. Lett. {\bf 123}, 110602 (2019).
\bibitem{LGU2020} K. Liu, Z. Gong, and M. Ueda, Thermodynamic Uncertainty Relation for Arbitrary Initial States, Phys. Rev. Lett. {\bf 125},140602 (2020).
\bibitem{KS2020} T. Koyuk and U. Seifert, Thermodynamic Uncertainty Relation for Time-Dependent Driving, Phys. Rev. Lett. {\bf 125}, 260604 (2020).
\bibitem{BS2016} A. C. Barato and U. Seifert, Cost and precision of Brownian clocks, Phys. Rev. X {\bf 6}, 041053 (2016).
\bibitem{NSB2018} B. Nguyen, U. Seifert, and A. C. Barato, Phase transition in thermodynamically consistent biochemical oscillators, J. Chem. Phys. {\bf 149}, 045101 (2018).
\bibitem{CL2024} Y. Cao and S. Liang, Stochastic thermodynamics for biological functions, Quant. Biol. {\bf 13}, e75 (2025).
\bibitem{MCH2019} R. Marsland III, W. Cui, and J. M. Horowitz, The thermodynamic uncertainty relation in biochemical oscillations, J. R. Soc. Interface {\bf 16}, 20190098 (2019).
\bibitem{YI2021} K. Yoshimura and S. Ito, Thermodynamic uncertainty relation and thermodynamic speed limit in deterministic chemical reaction networks, Phys. Rev. Lett. {\bf 127}, 160601 (2021).
\bibitem{KH2021} P. Kim and C. Hyeon. Thermodynamic optimality of glycolytic oscillations, J. Phys. Chem. B {\bf 125}, 5740 (2021).
\bibitem{CWQT2015} Y. Cao, H. Wang, Q. Ouyang, and Y. Tu, The free-energy cost of accurate biochemical oscillations, Nat. Phys. {\bf 11}, 772 (2015).
\bibitem{FCQT2018} C. Fei, Y. Cao, Q. Ouyang, and Y. Tu, Design principles for enhancing phase sensitivity and suppressing phase fluctuations simultaneously in biochemical oscillatory systems, Nat. Commun. {\bf 9}, 1434 (2018).
\bibitem{CJH2020} Z. Cao, H. Jiang, and Z. Hou, Design principles for biochemical oscillations with limited energy resources, Phys. Rev. Research {\bf 2}, 043331 (2020).
\bibitem{OSB2022} L. Oberreiter, U. Seifert, and A. C. Barato, Universal minimal cost of coherent biochemical oscillations, Phys. Rev. E {\bf 106}, 014106 (2022).
\bibitem{SF2025} D. Santolin and G. Falasco, Dissipation Bounds the Coherence of Stochastic Limit Cycles, Phys. Rev. Lett. {\bf 135}, 057101 (2025).
\bibitem{K2025} A. Kolchinsky, Comment on ``Dissipation bounds the coherence of stochastic limit cycles", arXiv:2510.14101v3.
\bibitem{NI2025} R. Nagayama and S. Ito, Duality between dissipation-coherence trade-off and thermodynamic speed limit based on thermodynamic uncertainty relation for stochastic limit cycles, arXiv:2509.06421v3.
\bibitem{FCS2020} L. P. Fischer, H.-M. Chun, and U. Seifert, Free diffusion bounds the precision of currents in underdamped dynamics, Phys. Rev. E {\bf 102}, 012120 (2020).
\bibitem{CFS2019} H.-M. Chun, L. P. Fischer, and U. Seifert, Effect of a magnetic field on the thermodynamic uncertainty relation, Rev. E {\bf 99}, 042128 (2019).
\bibitem{P2022} P. Pietzonka, Classical Pendulum Clocks Break the Thermodynamic Uncertainty Relation, Phys. Rev. Lett. {\bf 128}, 130606 (2022).
\bibitem{GEF2024} A. Gopal, M. Esposito, and N. Freitas, Thermodynamic cost of precise timekeeping in an electronic underdamped clock, Phys. Rev. B {\bf 109}, 085421 (2024).
\bibitem{SVH2024} D. Scheer, J. V\"{o}ller, and F. Hassler, The superconducting clock-circuit: Improving the coherence of Josephson radiation beyond the thermodynamic uncertainty relation, SciPost Phys. {\bf 17}, 140 (2024).
\bibitem{CP2026} E. P. Cital and V. Holubec, Inertia tames fluctuations in autonomous stationary heat engines, New J. Phys. {\bf 28}, 034605 (2026).
\bibitem{CP2026_2} E. P. Cital and V. Holubec, Strong violation of the thermodynamic uncertainty relation in a minimal autonomous heat engine, Phys. Rev. E {\bf 114}, 024153 (2026).
\bibitem{VH2019} T. Van Vu and Y. Hasegawa, Uncertainty relations for underdamped Langevin dynamics, Phys. Rev. E {\bf 100}, 032130 (2019).
\bibitem{LPP2019} J. S. Lee, J.-M. Park, and H. Park, Thermodynamic uncertainty relation for underdamped Langevin systems driven by a velocity-dependent force, Phys. Rev. E {\bf 100}, 062132 (2019).
\bibitem{LPP2021} J. S. Lee, J.-M. Park, and H. Park, Universal form of thermodynamic uncertainty relation for Langevin dynamics, Phys. Rev. E {\bf 104}, L052102 (2021).
\bibitem{D2022} A. Dechant, Bounds on the precision of currents in underdamped Langevin dynamics, arXiv:2202.10696v1.
\bibitem{Stz2001} S. H. Strogatz, {\it Nonlinear Dynamics and Chaos: With Applications to Physics, Biology, Chemistry, and Engineering} (Westview Press, Boulder, CO, 2001).
\bibitem{DX2013} R. Du and L. Xie, {\it The Mechanics of Mechanical Watches and Clocks} (Springer, Berlin, 2013).
\bibitem{SF2012} R. E. Spinney and I. J. Ford, Entropy production in full phase space for continuous stochastic dynamics, Phys. Rev. E {\bf 85}, 051113 (2012).
\bibitem{SM} Supplemental Material at [inserted by publisher].
\bibitem{KQ2004} K. H. Kim and H. Qian, Entropy Production of Brownian Macromolecules with Inertia, Rhys. Rev. Lett. {\bf 93}, 120602 (2004).
\bibitem{HSE2020} T. Herpich, K. Shayanfard, and M. Esposito, Effective thermodynamics of two interacting underdamped Brownian particles, Phys. Rev. E {\bf 101}, 022116 (2020).
\bibitem{ZY1987} W. Q. Zhu and J. S. Yu, On the response of the Van Der Pol oscillator to white noise excitation, J. Sound Vib. {\bf 117}, 421 (1987).
\bibitem{H2014} P. Hoyng, Dynamics and performance of clock pendulums, Am. J. Phys. {\bf 82}, 1053 (2014).
\bibitem{Z2021} D. Ziemkiewicz, Numerical analysis of grasshopper escapement, Phys. Rev. E {\bf 103}, 062208 (2021).
\bibitem{Z2022} D. Ziemkiewicz, Entropy of timekeeping in a mechanical clock, Phys. Rev. E {\bf 105}, 055001 (2022).
\bibitem{CS2014} M. V. Chubynsky and G. W. Slater, Diffusing Diffusivity: A Model for Anomalous, yet Brownian, Diffusion, Phys. Rev. Lett. {\bf 113}, 098302 (2014).
\bibitem{CSMS2017} A. V. Chechkin, F. Seno, R. Metzler, and I. M. Sokolov, Brownian yet Non-Gaussian Diffusion: From Superstatistics to Subordination of Diffusing Diffusivities, Phys. Rev. X {\bf 7}, 021002 (2017).
\bibitem{K1984} Y. Kuramoto, {\it Chemical Oscillations, Waves, and Turbulence} (Springer, New York, 1984).
\bibitem{PRK2001} A. Pikovsky, M. Rosenblum, and J. Kurths, {\it Synchronization: A Universal Concept in Nonlinear Sciences} (Cambridge University Press, Cambridge, 2001).
\bibitem{IKS2016} Y. Izumida, H. Kori, and U. Seifert, Energetics of synchronization in coupled oscillators rotating on circular trajectories, Phys. Rev. E {\bf 94}, 052221 (2016).
\bibitem{LHJ2018} S. Lee, C. Hyeon, and J. Jo, Thermodynamic uncertainty relation of interacting oscillators in synchrony, Phys. Rev. E {\bf 98}, 032119 (2018).
\bibitem{CEP2026} M. Chudak, M. Esposito, and K. Ptaszy\ifmmode \acute{n}\else \'{n}\fi{}ski, Synchronization of thermodynamically consistent stochastic phase oscillators, Phys. Rev. E {\bf 113}, 034129 (2026).


\end{references}

\begin{references}
\bibitem{US2025_S} U. Seifert, {\it Stochastic Thermodynamics} (Cambridge University Press, Cambridge, 2025).
\end{references}
\end{document}